\documentclass[conference]{IEEEtran}
\IEEEoverridecommandlockouts

\usepackage{cite}
\usepackage{amsmath,amssymb,amsfonts}
\usepackage{algorithmic}
\usepackage{graphicx}
\usepackage{textcomp}
\usepackage{xcolor}
\usepackage{multirow,siunitx}
\usepackage[belowskip=-5pt,aboveskip=5pt]{caption}

\makeatletter

\def\ps@headings{%
\def\@oddhead{\parbox[t][\height][t]{\textwidth}{\centering
2018 IEEE 18th International Conference on Data Mining - IWSC'18 2nd International Workshop on Social Computing\\
}\hfil\hbox{}}%

\def\@evenhead{\parbox[t][\height][t]{\textwidth}{\centering
14-th IEEE International Conference on Peer-to-Peer Computing\\
}\hfil\hbox{}}%

}

\def\ps@IEEEtitlepagestyle{%
\def\@oddhead{\parbox[t][\height][t]{\textwidth}{\centering
 2018 IEEE 18th International Conference on Data Mining - IWSC'18 2nd International Workshop on Social Computing\\
}\hfil\hbox{}}%
\def\@evenhead{\scriptsize\thepage \hfil \leftmark\mbox{}}%
}

\makeatother
\def\BibTeX{{\rm B\kern-.05em{\sc i\kern-.025em b}\kern-.08em
    T\kern-.1667em\lower.7ex\hbox{E}\kern-.125emX}}
\begin{document}
\title{Location, Occupation, and Semantics based Socioeconomic Status Inference on Twitter
\thanks{This work was supported by the SoSweet ANR project (ANR-15-CE38-0011) and the ACADEMICS project financed by IDEX LYON.}
}

\author{\IEEEauthorblockN{%
Jacob Levy Abitbol\IEEEauthorrefmark{1},
M\'arton Karsai\IEEEauthorrefmark{1}, and
Eric Fleury\IEEEauthorrefmark{2}}
\IEEEauthorblockA{\IEEEauthorrefmark{1}Univ Lyon, Inria, CNRS, ENS de Lyon, Universit\'e Claude Bernard Lyon 1, LIP UMR 5668, F-69007 LYON, France \\ \IEEEauthorrefmark{2}Inria, F-75012 Paris, France}
\IEEEauthorblockA{jacob.levy-abitbol@ens-lyon.fr, marton.karsai@ens-lyon.fr, eric.fleury@inria.fr}}


\maketitle

\begin{abstract}
The socioeconomic status of people depends on a combination of individual characteristics and environmental variables, thus its inference from online behavioral data is a difficult task. Attributes like user semantics in communication, habitat, occupation, or social network are all known to be determinant predictors of this feature. In this paper we propose three different data collection and combination methods to first estimate and, in turn, infer the socioeconomic status of French Twitter users  from their online semantics. Our methods are based on open census data, crawled professional profiles, and remotely sensed, expert annotated information on living environment. Our inference models reach similar performance of earlier results with the advantage of relying on broadly available datasets and of providing a generalizable framework to estimate socioeconomic status of large numbers of Twitter users. These results may contribute to the scientific discussion on social stratification and inequalities, and may fuel several applications.
\end{abstract}


\section{Introduction}

Online social networks have become one of the most disruptive communication platforms, as everyday billions of individuals use them to interact with each other. Their penetration in our everyday lives seems ever-growing and has in turn generated a massive volume of publicly available data open to analysis. The digital footprints left across these multiple media platforms provide us with a unique source to study and understand how the linguistic phenotype of a given user is related to social attributes such as socioeconomic status (SES).

The quantification and inference of SES of individuals is a long lasting question in the social sciences. It is a rather difficult problem as it may depend on a combination of individual characteristics and environmental variables~\cite{liere1980social}. Some of these features can be easier to assess like income, gender, or age  whereas others, relying to some degree on self-definition and sometimes entangled with privacy issues, are harder to assign like  ethnicity, occupation, education level or home location.
Furthermore, individual SES correlates with other individual or network attributes, as users tend to build social links with others of similar SES, a phenomenon known as status homophily\cite{mcpherson2001birds}, arguably driving  the observed stratification of society~\cite{leo2016socioeconomic}. At the same time, shared social environment, similar education level, and social influence have been shown to jointly lead socioeconomic groups to exhibit stereotypical behavioral patterns, such as shared political opinion~\cite{brown2017politics} or similar linguistic patterns~\cite{abitbol2018socioeconomic}. Although these features are entangled and causal relation between them is far from understood, they appear as correlations in the data.

Datasets recording multiple characteristics of human behaviour are more and more available due to recent developments in data collection technologies and increasingly popular online platforms and personal digital devices. The automatic tracking of online activities, commonly associated with profile data and meta-information; the precise recording of daily activities, interaction dynamics and  mobility patterns collected through mobile personal devices; together with the detailed and expert annotated census data all provide new grounds for the inference of individual features or behavioral patterns~\cite{lazer2009life}. The exploitation of these data sources has already been proven to be fruitful as cutting edge recommendation systems, advanced methods for health record analysis, or successful prediction tools for social behaviour heavily rely on them~\cite{kosinski2013private}. Nevertheless, despite the available data, some inference tasks, like individual SES prediction, remain an open challenge.

The precise inference of SES would contribute to overcome several scientific challenges and could potentially have several commercial applications~\cite{leo2018correlations}. Further,  robust SES inference would provide unique opportunities to gain deeper insights on socioeconomic inequalities~\cite{piketty2014capital}, social stratification~\cite{leo2016socioeconomic}, and on the driving mechanisms of network evolution, such as status homophily or social segregation.

In this work, we take a horizontal approach to this problem and explore various ways to infer the SES of a large sample of social media users. We propose different data collection and combination strategies using open, crawlable, or expert annotated socioeconomic data for the prediction task. Specifically, we use an extensive Twitter dataset of 1.3M users located in France, all associated with their tweets and profile information; 32,053 of them having inferred home locations. Individual SES is estimated by relying on three separate datasets, namely socioeconomic census data; crawled profession information and expert annotated Google Street View images of users' home locations. Each of these datasets is then used as ground-truth to infer the SES of Twitter users from profile and semantic features similar to~\cite{lampos2015}. We aim to explore and assess how the SES of social media users can be obtained and how much the inference problem depends on annotation and the  user's individual and linguistic attributes.

We provide in Section~\ref{sec:relwork} an overview of the related literature to contextualize the novelty of our work. In Section~\ref{sec:data} we provide a detailed description of the data collection and combination methods. In Section~\ref{sec:features} we introduce the  features extracted to solve the SES inference problem, with results summarized in Section~\ref{sec:results}. Finally, in Section~\ref{sec:limits} and~\ref{sec:concl} we conclude our paper with a brief discussion of the limitations and perspectives of our methods.

\section{Related works}
\label{sec:relwork}

There is a growing effort in the field to combine online behavioral data with census records, and expert annotated information to infer social attributes of users of online services. The predicted attributes range from easily assessable individual characteristics such as age~\cite{chamberlain}, or occupation~\cite{lampos2015,lampos2016,preot2015,tianran_hu} to more complex psychological and sociological traits like political affiliation~\cite{volkova}, personality~\cite{schwartz}, or SES~\cite{luo2017inferring,lampos2015}.

Predictive features proposed to infer the desired attributes are also numerous. In case of Twitter, user  information can be publicly queried within the limits of the public API~\cite{twitterAPI}. User characteristics collected in this way, such as  profile features, tweeting behavior, social network and linguistic content have been used for prediction, while other inference methods relying on external data sources such as website traffic data~\cite{culotta} or census data~\cite{EMoro,eisenstein} have also proven effective. Nonetheless, only recent works involve user semantics in a broader context related to social networks, spatiotemporal information, and personal attributes~\cite{preot2015, lampos2015, lampos2016, aletras2018predicting}.

The tradition of relating SES of individuals to their language dates back to the early stages of sociolinguistics where it was first shown that social status reflected through a person's occupation is a determinant factor in the way language is used~\cite{bernstein}. This line of research was recently revisited by Lampos \emph{et al.} to study the SES inference problem on Twitter. In a series of works~\cite{preot2015, lampos2015, lampos2016, aletras2018predicting}, the authors applied Gaussian Processes to predict user income, occupation and socioeconomic class based on demographic, psycho-linguistic features and a standardized job classification taxonomy which mapped Twitter users to their professional occupations.  The high predictive performance has proven this concept with $r=0.633$ for income prediction, and a precision of $55\%$ for 9-ways SOC classification, and $82\%$ for binary SES classification. Nevertheless, the models developed by the authors are learned by relying on datasets, which were manually labeled through an annotation process crowdsourced through Amazon Mechanical Turk at a high monetary cost. Although the labeled data has been released and provides the base for new extensions ~\cite{chamberlain}, it has two potential shortfalls that need to be acknowledged. First, the method requires access to a detailed job taxonomy, in this case specific to England, which hinders potential extensions of this line of work to other languages and countries. Furthermore, the language to income pipeline seems to show some dependency on the sample of users that actively chose to disclose their profession in their Twitter profile. Features obtained on this set might not be easily recovered from a wider sample of Twitter users. This limits the generalization of these results without assuming a costly acquisition of a new dataset.

\section{Data collection and combination}
\label{sec:data}

Our first motivation in this study was to overcome earlier limitations by exploring alternative data collection and combination methods. We provide here three ways to estimate the SES of Twitter users by using (a) open census data, (b) crawled and manually annotated data on professional skills and occupation, and (c) expert annotated data on home location Street View images. We provide here a collection of procedures that enable interested researchers to introduce predictive performance and scalability considerations when interested in developing language to SES inference pipelines.
In the following we present in detail all of our data collection and combination methods.

\subsection{Twitter corpus}

Our central dataset was collected from Twitter, an online news and social networking service. Through Twitter, users can post and interact by ``tweeting" messages with restricted length. Tweets may come with several types of metadata including information about the author's profile, the detected language as well as where and when the tweet was posted. Specifically, we recorded 90,369,215 tweets written in French, posted by 1.3 Million users in the timezones GMT and GMT+1 over one year (between August 2014 to July 2015)~\cite{TwitterDataEthic}. These tweets were obtained via the Twitter Powertrack API provided by Datasift with an access rate of $15\%$. Using this dataset we built several other corpora:

\subsubsection{Geolocated users}
\label{sec:geousers}

To find users with a representative home location we followed the method published in~\cite{allen,hu2016}. As a bottom line, we concentrated on $127,614$ users who posted at least five geolocated tweets with valid GPS coordinates, with at least three of them within a valid census cell (for definition see later), and over a longer period than seven days. Applying these filters we obtained 1,000,064 locations from geolocated tweets. By focusing on the geolocated users, we kept those with limited mobility, i.e., with median distance between locations not greater than 30 \textit{km}, with tweets posted at places and times which did not require travel faster than 130 $km/h$ (maximum speed allowed within France), and with no more than three tweets within a two seconds window. We further filtered out tweets with coordinates corresponding to locations referring to places (such as ``Paris" or ``France"). Thus, we removed locations that didn't  exactly correspond to GPS-tagged tweets and also users which were most likely bots. Home location was estimated by the most frequent location for a user among all coordinates he visited. This way we obtained $32,053$ users, each associated with a unique home location. Finally, we collected the latest $3,200$ tweets from the timeline of all of geolocated users using the Twitter public API~\cite{twitterAPI}. Note, that by applying these consecutive filters we obtained a more representative population as the Gini index, indicating overall socioeconomic inequalities, was $37.3\%$ before filtering become $36.4\%$ due to the filtering methods, which is closer to the value reported by the World Bank ($33.7\%$)~\cite{GiniWB}.

\begin{figure}[h]
\centerline{\includegraphics[width=0.8\columnwidth]{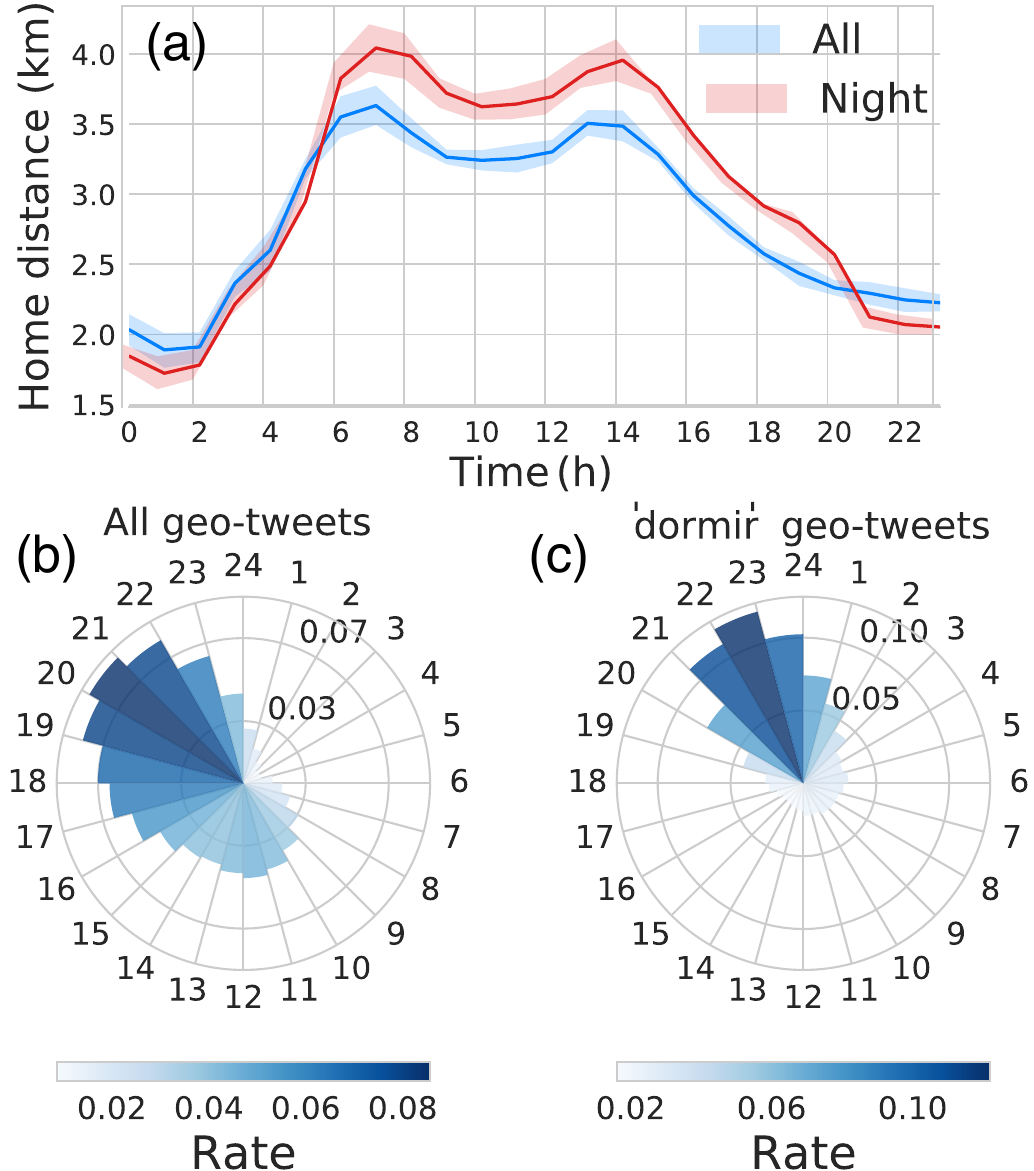}}
\caption{(a) Average distance from home of active users per hour of the day. (b) Hourly rate of all geolocated tweets and (c) geolocated tweets mentioning `dormir' averaged over all weekdays.}
\label{fig:home}
\end{figure}
To verify our results, we computed the average weekly distance from each recorded location of a user to his inferred home location defined either as its most frequent location overall or among locations posted outside of work-hours from 9AM to 6PM (see Fig.~\ref{fig:home}a). This circadian pattern displays great similarity to earlier results~\cite{hu2016} with two maxima, roughly corresponding to times at the workplace, and a local minimum at 1PM due to people having lunch at home. We found that this circadian pattern was more consistent with earlier results~\cite{hu2016} when we considered all geolocated tweets (``All" in Fig.~\ref{fig:home}a) rather than only tweets including ``home-related" expressions (``Night" in Fig.~\ref{fig:home}a). To further verify the inferred home locations, for a subset of 29,389 users we looked for regular expressions in their tweets that were indicative of being at home~\cite{hu2016}, such as ``chez moi", ``bruit", ``dormir" or ``nuit". In Fig.~\ref{fig:home}c we show the temporal distribution of the rate of the word ``dormir" at the inferred home locations. This distribution appears with a peak around 10PM, which is very different from the overall distribution of geolocated tweets throughout the day considering any location (see Fig.~\ref{fig:home}b).

\subsubsection{Linguistic data}
\label{sec:lingdata}

To obtain meaningful linguistic data we pre-processed the incoming tweet streams in several ways. As our central question here deals with language semantics of individuals, re-tweets do not bring any additional information to our study, thus we removed them by default. We also removed any expressions considered to be semantically meaningless like URLs, emoticons, mentions of other users (denoted by the \texttt{@} symbol) and hashtags (denoted by the \texttt{\#} symbol) to simplify later post-processing. In addition, as a last step of textual pre-processing, we downcased and stripped the punctuation from the text of every tweet.

\subsection{Census data}
\label{sec:census}

\begin{figure}[htbp]
\centerline{\includegraphics[width=.9\columnwidth]{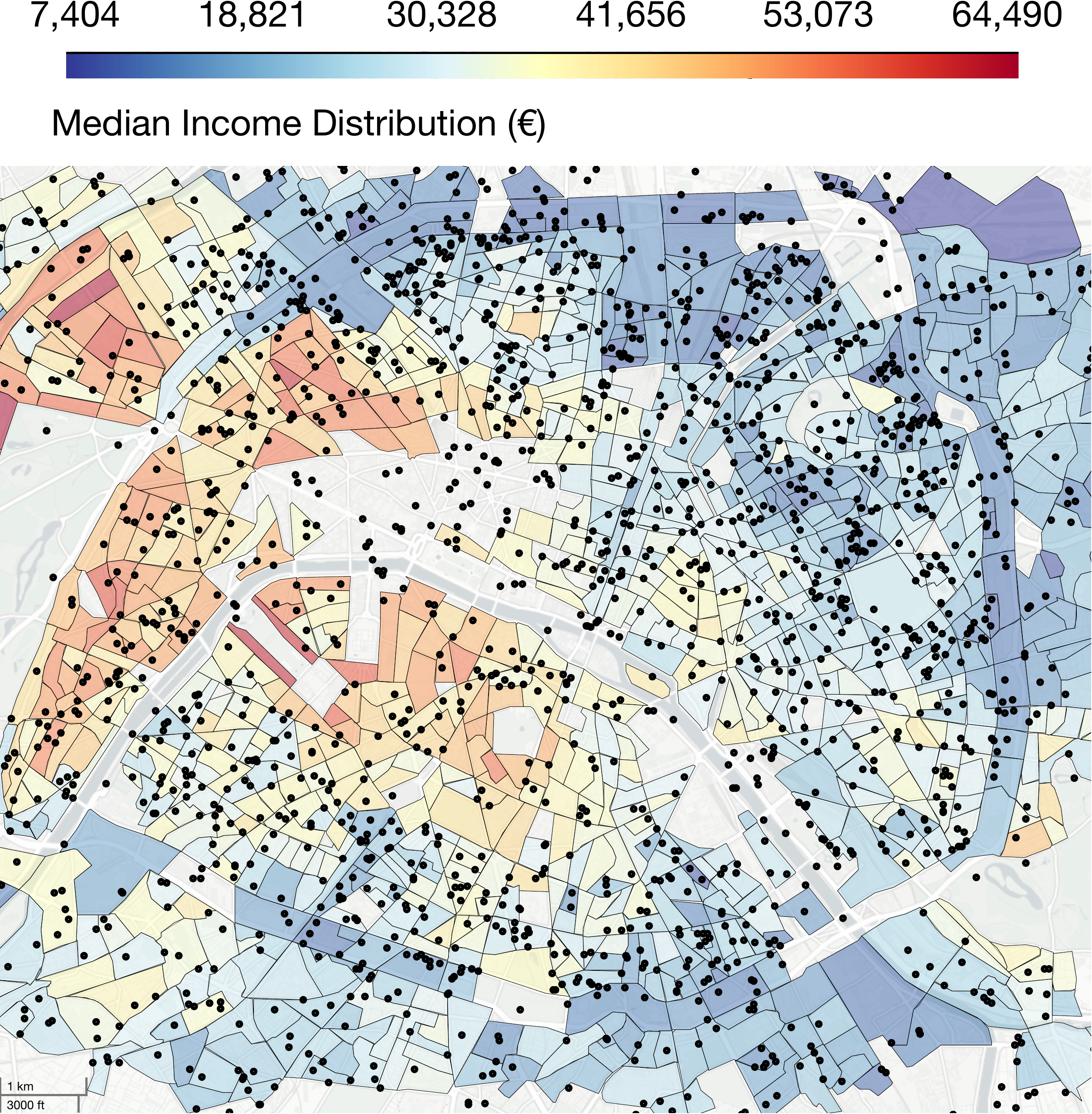}}
\caption{IRIS area cells in central Paris colored according to the median income of inhabitants, with inferred home locations of $2,000$ Twitter users.}
\label{fig:iris}
\end{figure}

Our first method to associate SES to geolocated users builds on an open census income dataset at intra-urban level for France~\cite{irisData}. Obtained from $2010$ French tax returns, it was released in December 2016 by the National Institute of Statistics and Economic Studies (INSEE) of France. This dataset collects detailed socioeconomic information of individuals at the census block level (called IRIS), which are defined as territorial cells with varying size but corresponding to blocks of around $2,000$ inhabitants, as shown in Fig.~\ref{fig:iris} for greater Paris. For each cell, the data records the deciles of the income distribution of inhabitants. Note that the IRIS data does not provide full coverage of the French territory, as some cells were not reported to avoid identification of individuals (in accordance with current privacy laws), or to avoid territorial cells of excessive area. Nevertheless, this limitation did not hinder our results significantly as we only considered users who posted at least three times from valid IRIS cells, as explained in Section~\ref{sec:geousers}.

To associate a single income value to each user, we identified the cell of their estimated home locations and assigned them with the median of the corresponding income distribution. Thus we obtained an average socioeconomic indicator for each user, which was distributed heterogeneously in accordance with Pareto's law~\cite{pareto1971manual}. This is demonstrated in Fig.~\ref{fig:SESSOC}a, where the $C(f)$ cumulative income distributions as the function of population fraction $f$ appears as a Lorentz-curve with area under the diagonal proportional to socioeconomic inequalities. As an example, Fig.~\ref{fig:iris} depicts the spatial distribution of $2,000$ users with inferred home locations in IRIS cells located in central Paris and colored as the median income.

\subsection{Occupation data}
\label{sec:occup}

Earlier studies~\cite{lampos2015,lampos2016,preot2015} demonstrated that annotated occupation information can be effectively used to derive precise income for individuals and infer therefore their SES. However, these methods required a somewhat selective set of Twitter users as well as an expensive annotation process by hiring premium annotators e.g. from Amazon Mechanical Turk. Our goal here was to obtain the occupations for a general set of Twitter users without the involvement of annotators, but by collecting data from parallel online services.

As a second method to estimate SES, we took a sample of Twitter users who mentioned their LinkedIn~\cite{LinkedIn} profile url in their tweets or Twitter profile. Using these pointers we collected professional profile descriptions from LinkedIn by relying on an automatic crawler mainly used in Search Engine Optimization (SEO) tasks~\cite{LinkedInSEO}. We obtained $4,140$ Twitter/LinkedIn users all associated with their job title, professional skills and profile description. Apart from the advantage of working with structured data, professional information extracted from LinkedIn is significantly more reliable than Twitter's due to the high degree of social scrutiny to which each profile is exposed~\cite{linkedin_priv}.

To associate income to Twitter users with LinkedIn profiles, we matched them with a given salary based on their reported profession and an occupational salary classification table provided by INSEE~\cite{INSEEsalary}. Due to the ambiguous naming of jobs and to acknowledge permanent/non-permanent, senior/junior contract types we followed three strategies for the matching. In $40\%$ of the cases we directly associated the reported job titles to regular expressions of an occupation. In $50\%$ of the cases we used string sequencing methods borrowed from DNA-sequencing~\cite{SeqMatch} to associate reported and official names of occupations with at least $90\%$ match. For the remaining $10\%$ of users we directly inspected profiles. The distribution of estimated salaries reflects the expected income heterogeneities as shown in Fig.~\ref{fig:SESSOC}. Users were eventually assigned to one of two SES classes based on whether their salary was higher or lower than the average value of the income distribution. Also note, that LinkedIn users may not be representative of the whole population. We discuss this and other types of poential biases in Section~\ref{sec:limits}.

\subsection{Expert annotated home location data}

Finally, motivated by recent remote sensing techniques, we sought to estimate SES via the analysis of the urban environment around the inferred home locations. Similar methodology has been lately reported by the remote sensing community~\cite{gebru2017} to predict socio-demographic features of a given neighborhood by analyzing Google Street View images to detect different car models, or to predict poverty rates across urban areas in Africa from satellite imagery ~\cite{jean_neal}. Driven by this line of work, we estimated the SES of geolocated Twitter users as follows:

\subsubsection{Pre-selection of home locations}
Using geolocated users identified in Section~\ref{sec:geousers}, we further filtered them to obtain a smaller set of users with more precise inferred home locations. We screened all of their geotagged tweets and looked for regular expressions determining whether or not a tweet was sent from home~\cite{hu2016}. As explained in Section~\ref{sec:geousers}, we exploited that ``home-suspected" expressions appeared with a particular temporal distribution (see Fig.~\ref{fig:home}c) since these expressions were used during the night when users are at home. This selection yielded $28,397$ users mentioning ``home-suspected" expressions regularly at their inferred home locations. 

\subsubsection{Identification of urban/residential areas}
\begin{figure}[htbp]
\centerline{\includegraphics[width=.9\columnwidth]{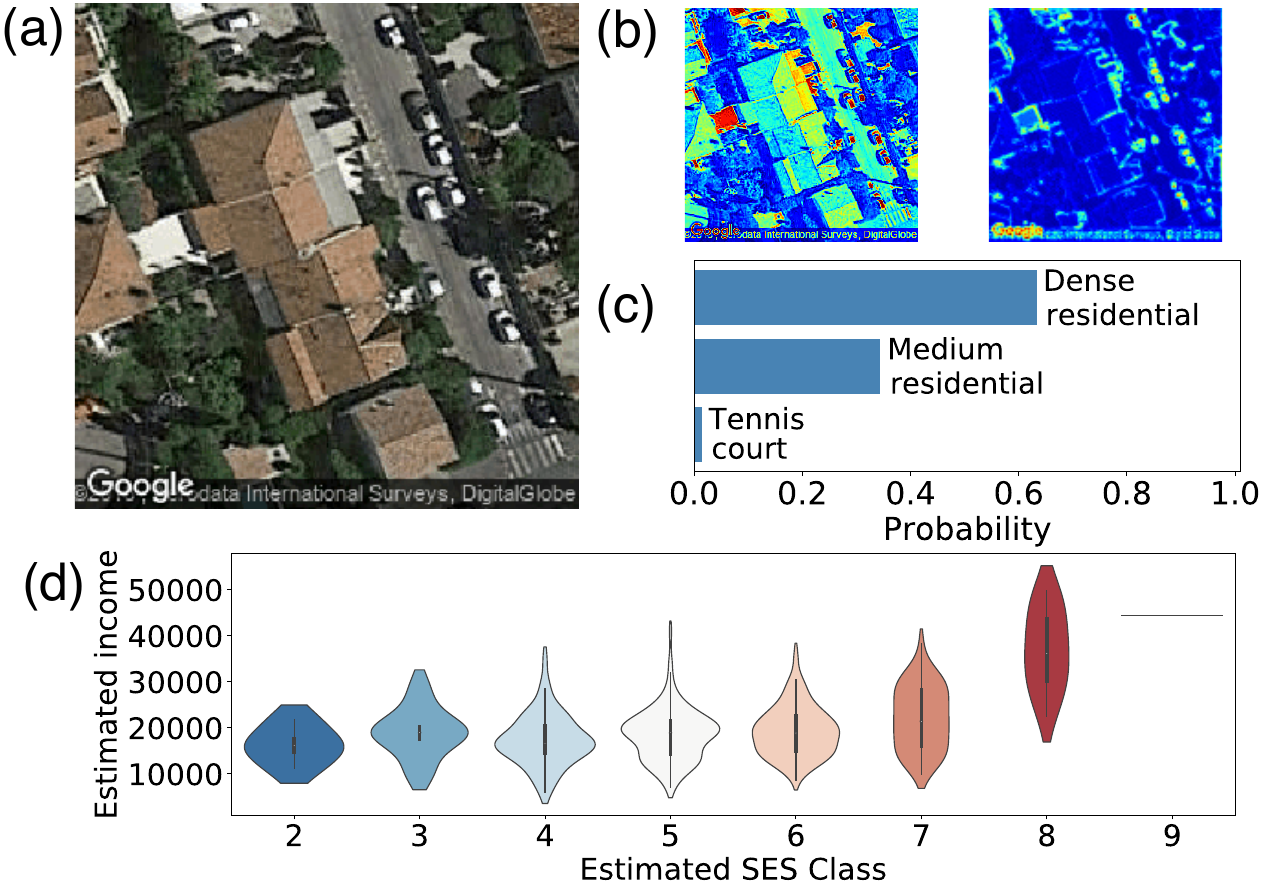}}
\caption{Top: ResNet50 Output: (a): Original satellite view; (b): First two hidden layers activation; (c): Final top-3 most frequent predicted area types; (d) Architect SES score agreement with census median income for the sampled home locations. It is shown as violin plots of income distributions for users annotated in different classes (shown on x-axis and by color).}
\label{fig:gmap}
\end{figure}

In order to filter out inferred home locations not in urban/residential areas, we downloaded via Google Maps Static API~\cite{gmapAPI} a satellite view in a $100m$ radius around each coordinate (for a sample see Fig.~\ref{fig:gmap}a). To discriminate between residential and non-residential areas, we built on land use classifier~\cite{castelluccio} using aerial imagery from the UC Merced dataset~\cite{ucMerced}. This dataset contains $2100$ $256\times 256$ $1m/px$ aerial RGB images over $21$ classes of different land use (for a pair of sample images see Fig.~\ref{fig:gmap}b). To classify land use a CaffeNet architecture was trained which reached an accuracy over $95\%$. Here, we instantiated a \emph{ResNet50} network using \emph{keras} ~\cite{chollet2015keras} pre-trained on ImageNet~\cite{imagenet_cvpr09} where all layers except the last five were frozen. The network was then trained with $10$-fold cross validation achieving a $93\%$ accuracy after the first $100$ epochs. We used this model to classify images of the estimated home location satellite views (cf. Figure~\ref{fig:gmap}a) and kept those which were identified as residential areas (see Fig.~\ref{fig:gmap}b, showing the activation of the two first hidden layers of the trained model). This way $5,396$ inferred home locations were discarded.

\subsubsection{Home location data with expert annotated SES}

Next we aimed to estimate SES from architectural/urban features associated to the home locations. Thus, for each home location we collected two additional satellite views at different resolutions as well as six Street View images, each with a horizontal view of approximately $ 90^{\circ}$. We randomly selected a sample of $1,000$ locations and involved architects to assign a SES score (from 1 to 9) to a sample set of selected locations based on the satellite and Street View around it (both samples had 333 overlapping locations). For validation, we took users from each annotated SES class and computed the distribution of their incomes inferred from the IRIS census data (see Section~\ref{sec:census}). Violin plots in Fig.~\ref{fig:gmap}d show that in expert annotated data, as expected, the inferred income values were positively correlated with the annotated SES classes. Labels were then categorized into two socioeconomic classes for comparison purposes. All in all, both annotators assigned the same label to the overlapping locations in $81.7\%$ of samples.

\begin{table}[h!]
\centering
\begin{tabular}{|p{1.5cm}|p{1.5cm}|p{1.5cm}|p{1.5cm}|}
\hline
 & \textbf{Census} & \textbf{Occupation} & \textbf{Expert} \\
\hline
\hline
 \textbf{Size} & $32,053$ & $4,140$ & $1,000$ \\ \hline
 \textbf{Low SES} & $0.54$ & $0.46$ & $0.58$  \\ \hline
 \textbf{High SES} & $0.46$ & $0.54$ & $0.42$ \\ 
\hline
\end{tabular}
\vspace{1em}
\caption{Number of users and estimated fractions of low and high SES in each dataset}
\label{tab:data}
\end{table}

\begin{figure}[htbp]
\centerline{\includegraphics[width=.95\columnwidth]{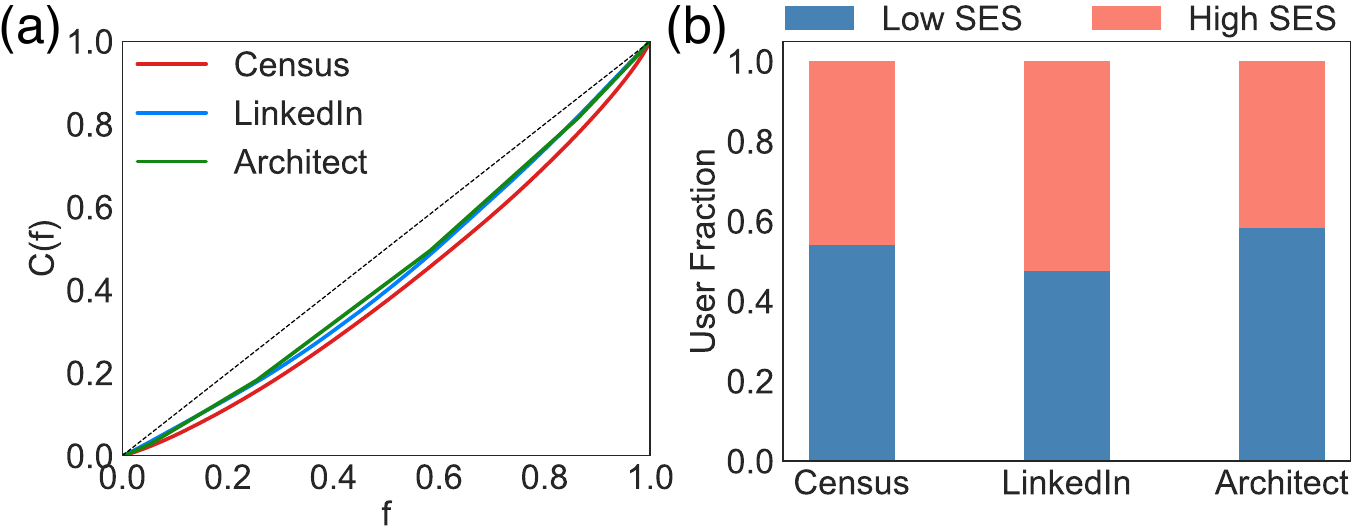}}
\caption{Cumulative distributions of income as a function of sorted fraction $f$ of individuals. Dashed line corresponds to the perfectly balanced distribution. Distributions appear similar in spite of dealing with heterogeneous samples.}
\label{fig:SESSOC}
\end{figure}

To solve the SES inference problem we used the above described three datasets (for a summary see Table~\ref{tab:data}). We defined the inference task as a two-way classification problem by dividing the user set of each dataset into two groups. For the census and occupation datasets the lower and higher SES classes were separated by the average income computed from the whole distribution, while in the case of the expert annotated data we assigned people from the lowest five SES labels to the lower SES class in the two-way task. The relative fractions of people assigned to the two classes are depicted in Fig.~\ref{fig:SESSOC}b for each dataset and summarized in Table~\ref{tab:data}.

\section{Feature selection}
\label{sec:features}

Using the user profile information and tweets collected from every account's timeline, we built a feature set for each user, similar to Lampos \emph{et al.}~\cite{lampos2015}. We categorized features into two sets, one containing shallow features directly observable from the data, while the other was obtained via a pipeline of data processing methods to capture semantic user features.

\subsection{User Level Features}
The user level features are based on the general user information or aggregated statistics about the tweets~\cite{lampos2016}. We therefore include general ordinal values such as the number and rate of retweets, mentions, and coarse-grained information about the social network of users (number of friends, followers, and ratio of friends to followers). Finally we vectorized each user's profile description and tweets and selected the top $450$ and $560$ 1-grams and 2-grams, respectively, observed through their accounts (where the rank of a given 1-gram was estimated via \textit{tf-idf}~\cite{leskovec2014mining}). 

\subsection{Linguistic features}

To represent textual information, in addition to word count data, we used topic models to encode coarse-grained information on the content of the tweets of a user, similar to~\cite{lampos2015}. This enabled us to easily interpret the relation between semantic and socioeconomic features. Specifically, we started by training a \textit{word2vec} model~\cite{mikolov2013efficient} on the whole set of tweets (obtained in the 2014-2015 timeframe) by using the skip-gram model and negative sampling with parameters similar to~\cite{lampos2016,chamberlain}. To scale up the analysis, the number of dimensions for the embedding was kept at $50$. This embedded words in the initial dataset in a $\mathbb{R}^{50}$ vector space.

\begin{figure}[htbp]
\centerline{\includegraphics[width=.95\columnwidth]{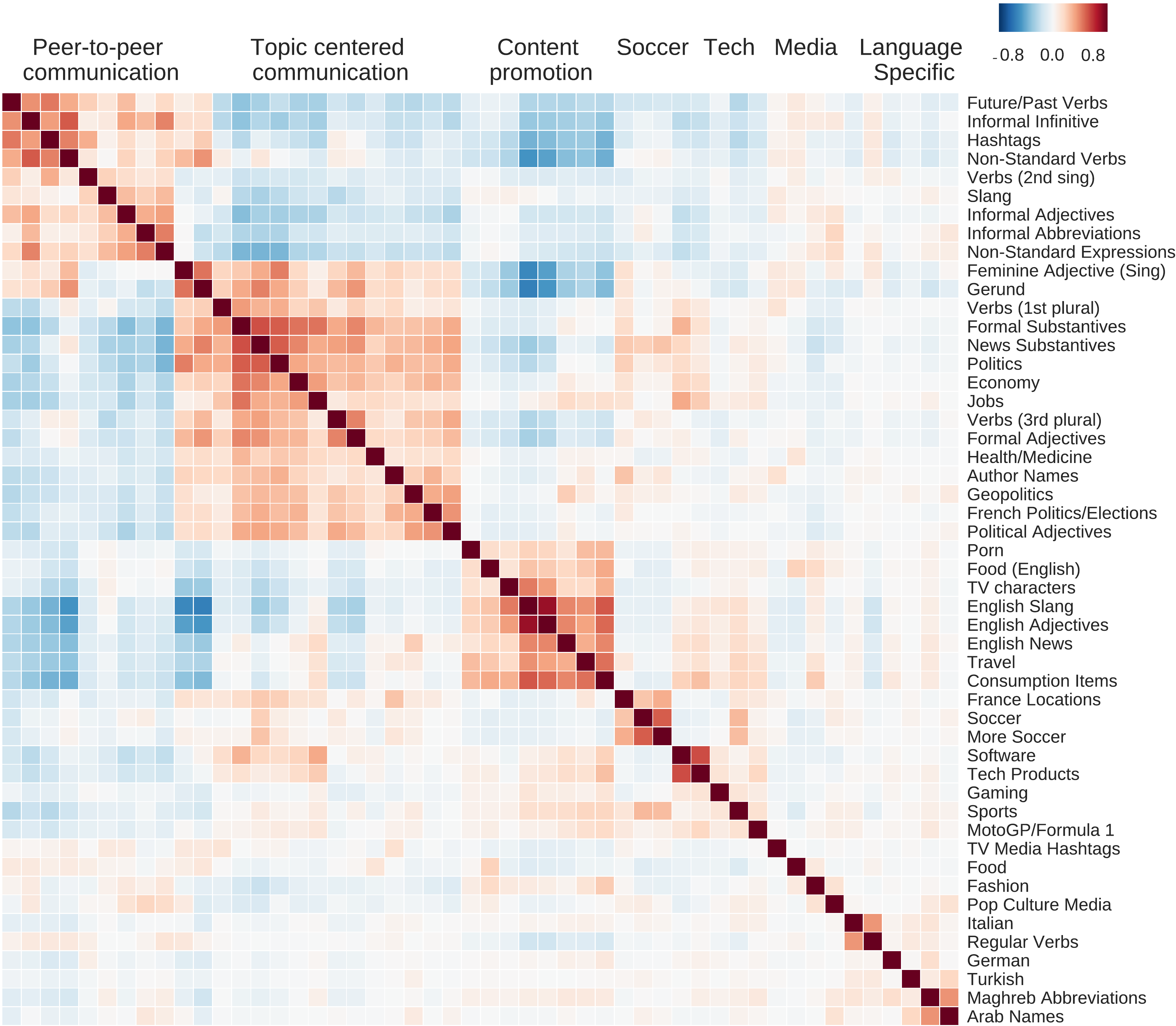}}
\caption{Clustered topic-to-topic correlation matrix: Topics are generated via the spectral clustering of the word2vec word co-similarity matrix. Row labels are the name of topics while column labels are their categories. Blue cells (resp. red) assign negative (resp. positive) Pearson’s correlation coefficients.}
\label{fig:topcorrs}
\end{figure}

Eventually we extracted conversation topics by running a spectral clustering algorithm on the word-to-word similarity matrix $M\in\mathbb{R}^{V\times V}$ with $V$ vocabulary size and elements defined as the $M_{ij}=\frac{\langle u_{i},u_{j}\rangle }{|| u_i || || u_j|| }$ cosine similarity between word vectors. Here $u_i\in \mathbb{R}^{50}$ is a vector of a word $i\in V$ in the embedding, $\langle \cdot \rangle$ is the dot product of vectors, and $||\cdot||$ is the $L^2$ norm of a vector. This definition allows for negative entries in the matrix to cluster, which were set to null in our case. This is consistent with the goal of the clustering procedure as negative similarities shouldn't encode dissimilarity between pairs of words but orthogonality between the embeddings. This procedure was run for $50$, $100$ and $200$ clusters and allowed the homogeneous distribution of words among clusters (hard clustering). The best results were obtained with $100$ topics in the topic model. Finally, we manually labeled topics based on the words assigned to them, and computed the topic-to-topic correlation matrix shown in Fig.~\ref{fig:topcorrs}. There, after block diagonalization, we found clearly correlated groups of topics which could be associated to larger topical areas such as communication, advertisement or  soccer.

As a result we could compute a representative topic distribution for each user, defined as a vector of normalized usage frequency of words from each topic. Also note that the topic distribution for a given user was automatically obtained as it depends only on the set of tweets and the learned topic clusters without further parametrization.

\begin{figure}[ht!]
\centerline{\includegraphics[width=.95\columnwidth]{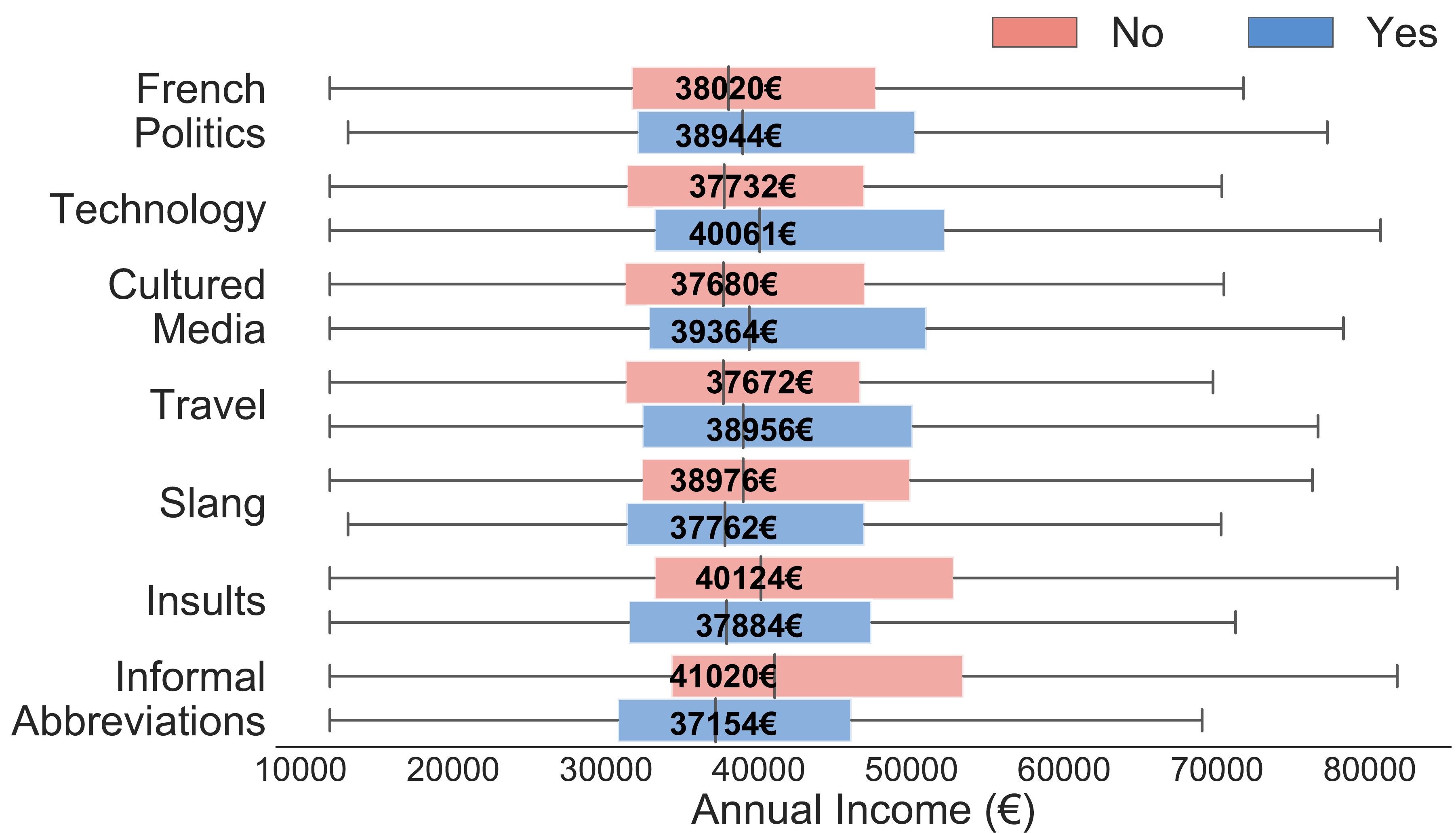}}
\caption{Average income for users who tweeted about a given topic (blue) vs. those who didn't (red). Label of the considered topic is on the left.}
\label{fig:topment}
\end{figure}

To demonstrate how discriminative the identified topics were in terms of the SES of users we associated to each user the 9th decile value of the income distribution corresponding to the census block of their home location and computed for each labelled topic the average income of users depending on whether or not they mentioned the given topic. Results in Fig. ~\ref{fig:topment} demonstrates that topics related to politics, technology or culture are more discussed by people with higher income, while other topics associated to slang, insults or informal abbreviations are more used by people of lower income. These observable differences between the average income of people, who use (or not) words from discriminative topics, demonstrates well the potential of word topic clustering used as features for the inference of SES. All in all, each user in our dataset was assigned with a $1117$ feature vector encoding the lexical and semantic profile she displayed on Twitter. We did not apply any further feature selection as the distribution of importance of features appeared rather smooth (not shown here). It did not provided evident ways to identify a clear set of particularly determinant features, but rather indicated that the combination of them were important.

\section{Results}
\label{sec:results}

In order to assess the degree to which linguistic features can be used for discriminating users by their socioeconomic class, we trained with these feature sets different learning algorithms. Namely, we used the XGBoost algorithm~\cite{xgboost}, an implementation of the gradient-boosted decision trees for this task. Training a decision tree learning algorithm involves the generation of a series of rules, split points or nodes ordered in a tree-like structure enabling the prediction of a target output value based on the values of the input features. More specifically, XGBoost, as an ensemble technique, is trained by sequentially adding a high number of individually weak but complementary classifiers to produce a robust estimator: each new model is built to be maximally correlated with the negative gradient of the loss function associated with the model assembly~\cite{torlay2017}. To evaluate the performance of this method we benchmarked it against more standard ensemble learning algorithms such as AdaBoost and Random Forest.
\begin{table}[h!]
\centering
\begin{tabular}{|p{2cm}|p{1.73cm}|p{1.73cm}|p{1.73cm}|}
\hline
 & \textbf{Census} & \textbf{Occupation} & \textbf{Expert} \\
\hline
\hline
 \textbf{AdaBoost} & $0.549 \pm 0.009$ & $0.628\pm0.022$ & $0.575 \pm0.013$ \\ \hline
 \textbf{Random Forest} & $0.677 \pm 0.011 $ & $0.783\pm0.017$ & $0.593\pm0.049$  \\ \hline
 \textbf{XGBoost} & $\mathbf{0.700 \pm 0.011}$ & $\mathbf{0.798\pm0.015}$ & $\mathbf{0.605\pm0.029}$ \\ 
\hline
\end{tabular}
\vspace{1em}
\caption{Classification performance (5-CV):  AUC scores (mean $\pm$ STD)  of three different classifiers on each dataset)}
\label{tab:dataAUC}
\end{table}

For each socioeconomic dataset, we trained our models by using 75\% of the available data for training and the remaining 25\% for testing. During the training phase, the training data undergoes a $k$-fold inner cross-validation, with $k=5$, where all splits are computed in a stratified manner to get the same ratio of lower to higher SES users. The four first blocks were used for inner training and the remainder for inner testing. This was repeated ten times for each model so that in the end, each model's performance on the validation set was averaged over $50$ samples. For each model, the parameters were fine-tuned by training 500 different models over the aforementioned splits. The selected one was that which gave the best performance on average, which was then applied to the held-out test set. This  is then repeated through a 5-fold outer cross-validation.

In terms of prediction score, we followed a standard procedure in the literature~\cite{auc} and evaluated the learned models by considering the area under the receiver operating characteristic curve (AUC). This metric can be thought as the probability that a classifier ranks a randomly chosen positive instance higher than a randomly chosen negative one\cite{torlay2017}. 

\begin{figure}[h]
\centering
\includegraphics[width=.8\columnwidth]{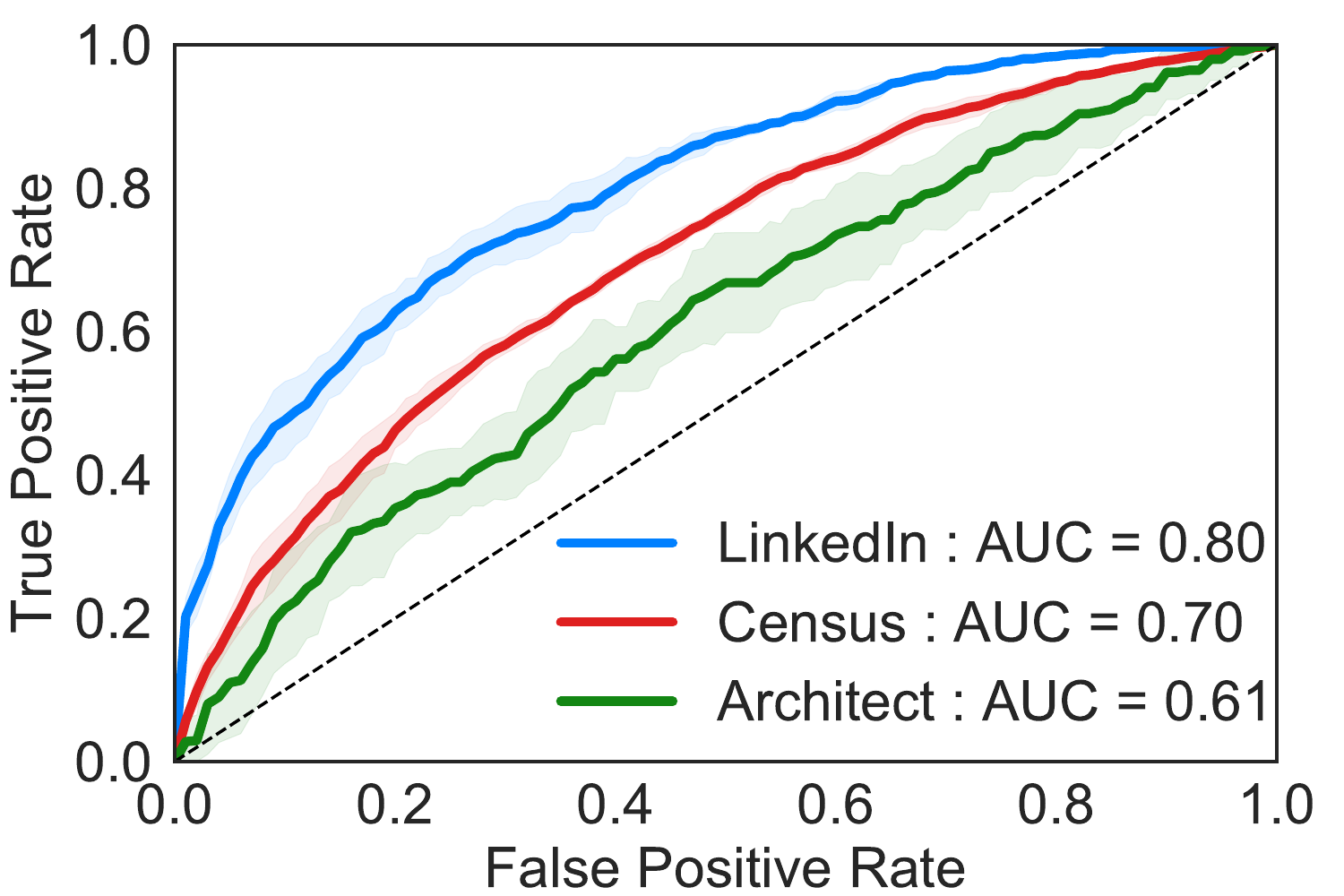}
\caption{ ROC curves for 2-way SES prediction using tuned XGBoost in each of the 3 SES datasets. AUC values are reported in the legend. The dashed line corresponds to the line of no discrimination. Solid lines assign average values over all folds while shaded regions represent standard deviation.}
\label{fig:SOC}
\end{figure}

This procedure was applied to each of our datasets. The obtained results are shown in Fig.~\ref{fig:SOC} and in Table~\ref{tab_res}.

As a result, we first observed that XGBoost consistently provided top prediction scores when compared to AdaBoost and Random Forest (all performance scores are summarised in Table~\ref{tab:dataAUC}). We hence used it for our predictions in the remainder of this study. We found that the LinkedIn data was the best, with $AUC=0.80$, to train a model to predict SES of people based on their semantic features. It provided a $10\%$ increase in performance as compared to the census based inference with $AUC=0.70$, and $19\%$ relative to expert annotated data with $AUC=0.61$. Thus we can conclude that there seem to be a trade-off between scalability and prediction quality, as while the occupation dataset provided the best results, it seems unlikely to be subject to any upscaling due to the high cost of obtaining a clean dataset. Relying on location to estimate SES seems to be more likely to benefit from such an approach, though at the cost of an increased number of mislabelled users in the dataset. Moreover, the annotator's estimation of SES using Street View at each home location seems to be hindered by the large variability of urban features. Note that even though inter-agreement is 76\%, the Cohen's kappa score for annotator inter-agreement is low at 0.169. Furthermore, we remark that the expert annotated pipeline was also subject to noise affecting the home location estimations, which potentially contributed to the lowest predictive performance. 

Finally, it should also be noted that following recent work by Aletras and Chamberlain in~\cite{aletras2018predicting}, we tested our model by extending the feature set with the \textit{node2vec} embedding of users computed from the mutual mention graph of Twitter. Nevertheless, in our setting,  it did not increase the overall predictive performance of the inference pipeline. We hence didn't include in the feature set for the sake of simplicity.

\begin{table}[h!]
\centering
    \begin{tabular}{|c|l|S|S|S|S|}           \hline 
\multirow{2}{*}{Dataset} &\multirow{2}{*}{SES Class}    & \multicolumn{3}{c|}{Performance on test set}\\    \cline{3-5}
 &  &{Precision} & {Recall}  &{F1-score}            \\   \hline 
\multirow{2}{*}{Census} & Low& 0.652 & 0.596 & 0.624  \\  
\cline{2-5}  
 & High & 0.628 & 0.682 & 0.652   \\   \hline 
 \multirow{2}{*}{LinkedIn} & Low& 0.700 & 0.733 & 0.717  \\   \cline{2-5}  
 & High & 0.735 & 0.702 & 0.720            \\   \hline 
 \multirow{2}{*}{Architect} & Low & 0.622  &  0.598 & 0.607  \\   \cline{2-5}  
 & High &  0.550   & 0.573   &   0.556 \\   \hline 
 \end{tabular}
 \vspace{1em}
 \caption{Detailed average performance (5-CV)  on test data for the binary SES inference problem for each of the 3 datasets}
 \label{tab_res}
 \end{table}

\section{Limitations}
\label{sec:limits}

In this work we combined multiple datasets collected from various sources. Each of them came with some bias due to the data collection and post-treatment methods or the incomplete set of users. These biases may limit the success of our inference, thus their identification is important for the interpretation and future developments of our framework.

$\bullet$ \emph{Location data}: Although we designed very strict conditions for the precise inference of home locations of geolocated users, this process may have some uncertainty due to outlier behaviour. Further bias may be induced by the relatively long time passed between the posting of the location data and of the tweets collection of users.

$\bullet$  \emph{Census data}: As we already mentioned the census data does not cover the entire French territory as it reports only cells with close to $2,000$ inhabitants. This may introduce biases in two ways: by limiting the number of people in our sample living in rural areas, and by associating income with large variation to each cell. While the former limit had marginal effects on our predictions, as Twitter users mostly live in urban areas, we addressed the latter effect by associating the median income to users located in a given cell.

$\bullet$  \emph{Occupation data}: LinkedIn as a professional online social network is predominantly used by people from IT, business, management, marketing or other expert areas, typically associated with higher education levels and higher salaries. Moreover, we could observe only users who shared their professional profiles on Twitter, which may further biased our training set. In terms of occupational-salary classification, the data in~\cite{INSEEsalary} was collected in $2010$ thus may not contain more recent professions. These biases may induce limits in the representativeness of our training data and thus in the predictions' precision. However, results based on this method of SES annotation performed best in our measurements, indicating that professions are among the most predictive features of SES, as has been reported in~\cite{lampos2015}.

$\bullet$  \emph{Annotated home locations}: The remote sensing annotation was done by experts and their evaluation was based on visual inspection and biased by some unavoidable subjectivity. Although their annotations were cross-referenced and found to be consistent, they still contained biases, like over-representative middle classes, which somewhat undermined the prediction task based on this dataset.

Despite these shortcomings, using all the three datasets we were able to infer SES with performances close to earlier reported results, which were based on more thoroughly annotated datasets. Our results, and our approach of using open, crawlable, or remotely sensed data highlights the potential of the proposed methodologies.

\section{Conclusions}
\label{sec:concl}

In this work we proposed a novel methodology for the inference of the SES of Twitter users. We built our models combining information obtained from numerous sources, including Twitter, census data, LinkedIn and Google Maps. We developed precise methods of home location inference from geolocation, novel annotation of remotely sensed images of living environments, and effective combination of datasets collected from multiple sources. As new scientific results, we demonstrated that within the French Twitter space, the utilization of words in different topic categories, identified via advanced semantic analysis of tweets, can discriminate between people of different income. More importantly, we presented a proof-of-concept that our methods are competitive in terms of SES inference when compared to other methods relying on domain specific information.

We can identify several future directions and applications of our work. First, further development of data annotation of remotely sensed information is a promising direction. Note that after training,  our model requires as input only information, which can be collected exclusively from Twitter, without relying on other data sources. This holds a large potential in terms of SES inference of larger sets of Twitter users, which in turn opens the door for studies to address population level correlations of SES with language, space, time, or the social network. This way our methodology has the merit not only to answer open scientific questions, but also to contribute to the development of new applications in recommendation systems, predicting customer behavior, or in online social services.

\section*{Acknowledgments}

We thank J-Ph. Magu\'e, J-P. Chevrot, D. Seddah, D. Carnino and E. De La Clergerie for constructive discussions and for their advice on data management and analysis. We are grateful to J. Altn\'eder and M. Hunyadi for their contributions as expert architects for data annotation.


\bibliographystyle{IEEEtran}

\begin{thebibliography}{10}
\bibitem{TwitterDataEthic}
The collection of this dataset was approved by the Ethic Committee of the
  hosting academic institute of the authors.
\end{thebibliography}

\end{document}